\RequirePackage{fix-cm}
\documentclass[twocolumn,epjc3]{svjour3}
\smartqed  
\RequirePackage{graphicx}
\RequirePackage[numbers,sort&compress]{natbib}
\RequirePackage[colorlinks,citecolor=blue,urlcolor=blue,linkcolor=blue]{hyperref}
\usepackage[T1]{fontenc}

\journalname{Eur. Phys. J. C}
\begin{document}

\title{Performance of a Large-Volume Cryogenic Pure CsI Detector for CE$\nu$NS and Low-Energy Rare-Event Searches
}

\titlerunning{Performance of a Large-Volume Cryogenic Pure CsI Detector}        

\author{Chen'guang Su\thanksref{addr1,e1,equal}
        \and
        Lingquan Kong\thanksref{addr1,e2,equal} 
        \and
        Qian Liu\thanksref{addr1,corresp} 
        \and
        Shi Chen\thanksref{addr1} 
        \and
        Wenqian Huang\thanksref{addr1} 
        \and
        Kimiya Moharrami\thanksref{addr1} 
        \and
        Yangheng Zheng\thanksref{addr1} 
        \and
        Jin Li\thanksref{addr1} 
}

\thankstext{e1}{e-mail: suchenguang17@mails.ucas.ac.cn}
\thankstext{e2}{e-mail: konglingquan19@mails.ucas.ac.cn}
\thankstext{equal}{These authors contributed equally to this work.}
\thankstext{corresp}{Corresponding author: \email{liuqian@ucas.ac.cn}}


\institute{School of Physical Sciences, University of Chinese Academy of Sciences, Beijing, 100049, China \label{addr1}
}

\date{Received: date / Accepted: date}

\maketitle

\begin{abstract}




A cryogenic detector system based on two 3.3 kg high-purity CsI crystals was developed and characterized at approximately 95 K. Each wedge-shaped crystal was coupled to dual-ended 3-inch photomultiplier tubes (PMTs) for scintillation readout. The measured light yields were $28.7 \pm 0.9$ and $29.3 \pm 1.0$ photoelectrons per keV electron-equivalent (PE/keV$_{ee}$) for the two crystals, with corresponding energy resolutions of 7.2 \% and 7.7 \% (FWHM) at 59.6 keV. The detector demonstrated excellent spatial uniformity, low intrinsic radioactivity, and stable operation over a continuous one-month period. Optical photon simulations using Geant4 reproduced the observed light collection trends, providing guidance for detector optimization. These results establish cryogenic pure CsI as a scalable technology for low-threshold rare-event searches.

\end{abstract}

\section{Introduction}
\label{intro}

Coherent elastic neutrino-nucleus scattering (CE$\nu$NS) is a Standard Model (SM) process first predicted in 1974 \cite{freedman1974coherent} and was first observed more than four decades later by the COHERENT collaboration in 2017 \cite{akimov2017COHERENT}, owing to the experimental challenges associated with detecting the extremely low-energy nuclear recoils it produces. In CE$\nu$NS, a low-energy neutrino scatters off an entire nucleus via $Z$-boson exchange, resulting in a minute nuclear recoil typically below 100 keV. The cross section is coherently enhanced by the square of the neutron number, making CE$\nu$NS a powerful probe for both nuclear structure and fundamental physics \cite{CEvNSFuture}.


The significance of CE$\nu$NS extends well beyond the confirmation of Standard Model predictions. It provides a sensitive probe of the weak mixing angle at low momentum transfer \cite{weakMixingAngle1,weakMixingAngle2ChargeRadius,weakMixingAngleNSILightMediator}, nuclear form factors \cite{formFactor1,formFactor2,formFactor3}, and neutrino electromagnetic properties such as the magnetic moment and charge radius \cite{magneticMomentNSI,magneticMoment,weakMixingAngle2ChargeRadius}. Furthermore, CE$\nu$NS offers a powerful channel to search for non-standard interactions (NSI) \cite{magneticMomentNSI,NSI2,NSI3,NSI4,NSI5,NSI6} and light mediator particles \cite{lightMediator,lightMediator2,weakMixingAngleNSILightMediator}, where the characteristic recoil spectrum is particularly sensitive to deviations at low energies. Since achieving lower detection thresholds directly enhances sensitivity to such effects, CE$\nu$NS experiments are uniquely positioned to explore new physics scenarios that manifest most strongly in the low-energy regime. At the same time, CE$\nu$NS constitutes an irreducible background for direct dark matter searches \cite{NeutrinoFog1,NeutrinoFog2}, highlighting its dual role in both precision measurements and exploratory searches for physics beyond the Standard Model \cite{DarkMatter3,DarkMatter4,DarkMatter5}.

However, the experimental observation of CE$\nu$NS and the realization of its full physics potential are constrained by the need for detectors with ultra-low energy thresholds, excellent energy resolution, and stable long-term performance \cite{lowerThreshold1,lowerThreshold2}. Achieving a low threshold is essential to increase the number of detectable CE$\nu$NS events, and this requirement is particularly stringent for reactor and solar neutrinos, whose spectra fall sharply at higher energies \cite{ReactorNeutrino,SolarNeutrino}. For these sources, sensitivity to the dominant low-energy region is only possible if the detection threshold is pushed sufficiently low. This capability not only maximizes the observable CE$\nu$NS signal but also strengthens searches for low-energy signatures of new physics, such as neutrino magnetic moments and light mediators. Meanwhile, excellent energy resolution is critical for reconstructing the recoil spectrum with fine granularity, allowing the detection of subtle spectral distortions that could indicate deviations from SM predictions, including NSI effects, form factor modifications, or electroweak precision parameters like the weak mixing angle \cite{SpectraDistortion1,SpectraDistortion2}. In this context, high light yield and good energy resolution are closely intertwined: a higher light yield not only reduces the detection threshold but also improves energy resolution by enhancing photon statistics. Nonetheless, optimizing light collection and reducing systematic contributions to energy resolution must be pursued simultaneously to fully exploit the shape sensitivity of CE$\nu$NS recoil spectra.

In this work, we present a detailed characterization of large-volume, cryogenic pure CsI crystals operated at approximately 95 K with dual-ended photomultiplier tube (PMT) readout. Despite their substantial mass (3.3 kg per crystal), the crystals achieved a light yield of 29 photoelectrons per keV electron-equivalent (PE/keV$_{ee}$ \cite{keVnr}) and an energy resolution better than 8 \% at 59.6 keV. Since the minimum detectable energy is determined by the smallest number of photoelectrons distinguishable from noise, a higher light yield directly translates into a lower energy threshold. The crystals further exhibit excellent spatial uniformity, low intrinsic radioactivity, and stable performance over a continuous operation period of one month. These results highlight cryogenic pure CsI as a compelling detector technology for future CE$\nu$NS experiments and other low-energy rare-event searches, including dark matter, where detection thresholds on the order of 1 keV${nr}$ are essential \cite{darkMatter1,darkMatter2}.

\section{Environmental setup}

\subsection{Detector module}

\begin{figure}
  \includegraphics[width=0.45\textwidth]{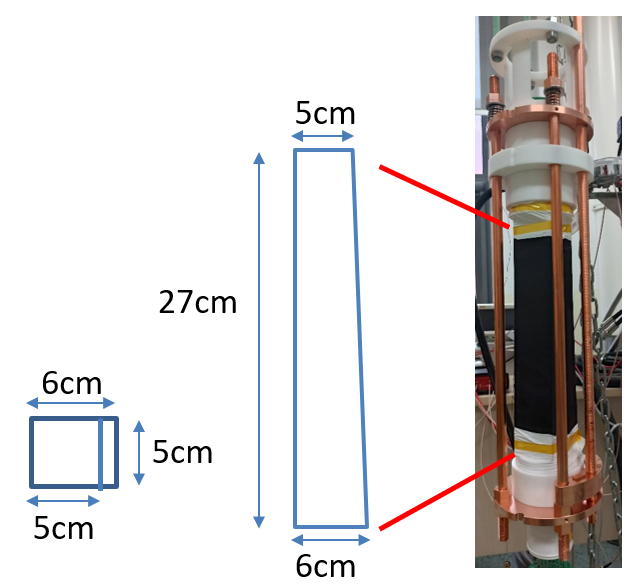}
\caption{Wedge-shaped crystal and copper assembly. The plot is not to scale. The black section is the crystal wrapped with Teflon and black paper. Both ends of the crystal are coupled to PMT, with the coupling areas wrapped in Teflon to prevent light leakage. }
\label{fig:crystalReal}
\end{figure}

The cryoCsI detector system comprises two high-purity CsI crystals, each with a mass of 3.3 kg, provided by the Shanghai Institute of Ceramics (SICAS). The two crystals, labeled NO.1 and NO.2, are housed within the modular detector assembly. Each crystal features a wedge-shaped geometry with a length of 27 cm and dual light output surfaces measuring 5 $\times$ 5 cm$^2$ and 5 $\times$ 6 cm$^2$ respectively, as shown in Fig.~\ref{fig:crystalReal}. All crystal surfaces are optically polished to enhance internal light transmission. The side surfaces are wrapped in four layers of Saint-Gobain Teflon Type BC642 to serve as reflective material and subsequently covered with black paper for optical isolation. Both ends of each crystal are directly coupled to 3-inch Hamamatsu R11065 photomultiplier tubes (PMTs). 

\begin{figure}
  \includegraphics[width=0.5\textwidth]{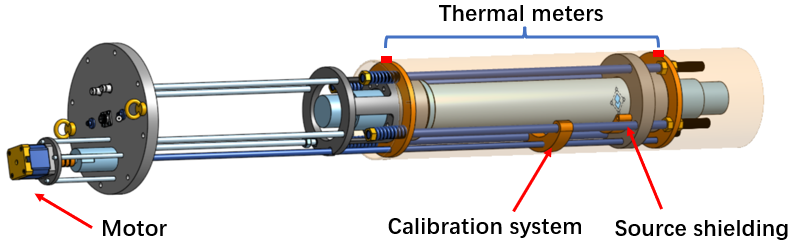}
\caption{Detector module: A single crystal with dual-end PMT coupling, a real-time collimated calibration source, and two thermal meters positioned at both ends. }
\label{fig:module}
\end{figure}

The support structure of the detector module is primarily constructed from copper and polyethylene. Four spring-loaded fixtures ensure uniform and robust optical coupling between the crystal surfaces and the PMTs. A built-in calibration system is integrated into the module, featuring a motorized mechanism that vertically translates a copper block holding a collimated $^{241}$Am source. This setup enables position-dependent energy calibration, as illustrated in Fig.~\ref{fig:module}. To monitor the thermal environment, two temperature sensors are mounted on the top and bottom copper plates in contact with the crystal ends.

The modular design offers flexibility for future upgrades, including the integration of larger or shape-optimized crystals, the replacement of PMTs with high-quantum-efficiency silicon photomultipliers (SiPMs) to further enhance light yield \cite{SiPM1,SiPM2}, and the implementation of alternative readout schemes.

\subsection{Cooling system}

\begin{figure}
  \includegraphics[width=0.48\textwidth]{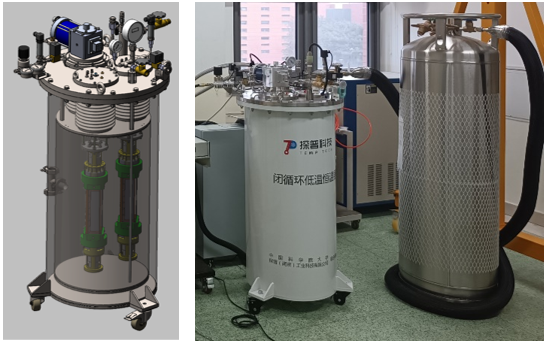}
\caption{Illustration of the dewar structure (left) and a photograph of the assembled dewar with the LN2 reservoir (right). }
\label{fig:dewar}
\end{figure}

The cooling system consists of a closed-cycle cryocooler dewar supplied by Temp Tech, supplemented by an external liquid nitrogen (LN2) reservoir, as shown in Fig.~\ref{fig:dewar}. The system maintains stable cryogenic performance at approximately 95 K. This working point was chosen because the PMTs do not operate reliably below 90 K. Although pure CsI generally exhibits slightly higher light yield closer to liquid-nitrogen temperature, measurements show that the variation between 70–100 K is modest \cite{LYTemp1Waveform,LYTemp2}. For the sake of PMT stability, this small compromise in light yield is acceptable. The temperature gradient between the two ends of the crystal remains within 2–3 K, a variation negligible in terms of its impact on light yield. The system typically requires around 24 hours to stabilize at the target temperature and can sustain cryogenic conditions continuously for years without LN2 replenishment.

All key components—including temperature regulation, gas pressure monitoring, PMT high-voltage control, data acquisition, and calibration source positioning—are integrated into remote control interfaces. This setup enables safe and full monitoring and operation without the need for physical access, which is particularly advantageous in radiation-controlled environments or during extended experimental runs.

\subsection{Shielding}


The dewar is enclosed within a fully 4$\pi$-covered, multi-layer passive shielding system composed of 5 cm of lead and 10 cm of high-density polyethylene (HDPE). 

In the present configuration, this shielding is used to suppress backgrounds such as environmental gamma rays, natural radioactivity, and cosmic rays, thereby enabling a precise assessment of the crystal’s intrinsic radiopurity. The outer lead layer primarily attenuates environmental gamma radiation, while the inner HDPE layer suppresses low-energy neutrons through elastic scattering and neutron capture. In future deployments near a neutrino source, the same shielding design will also mitigate source-related neutron backgrounds.

\begin{figure}
  \includegraphics[width=0.5\textwidth]{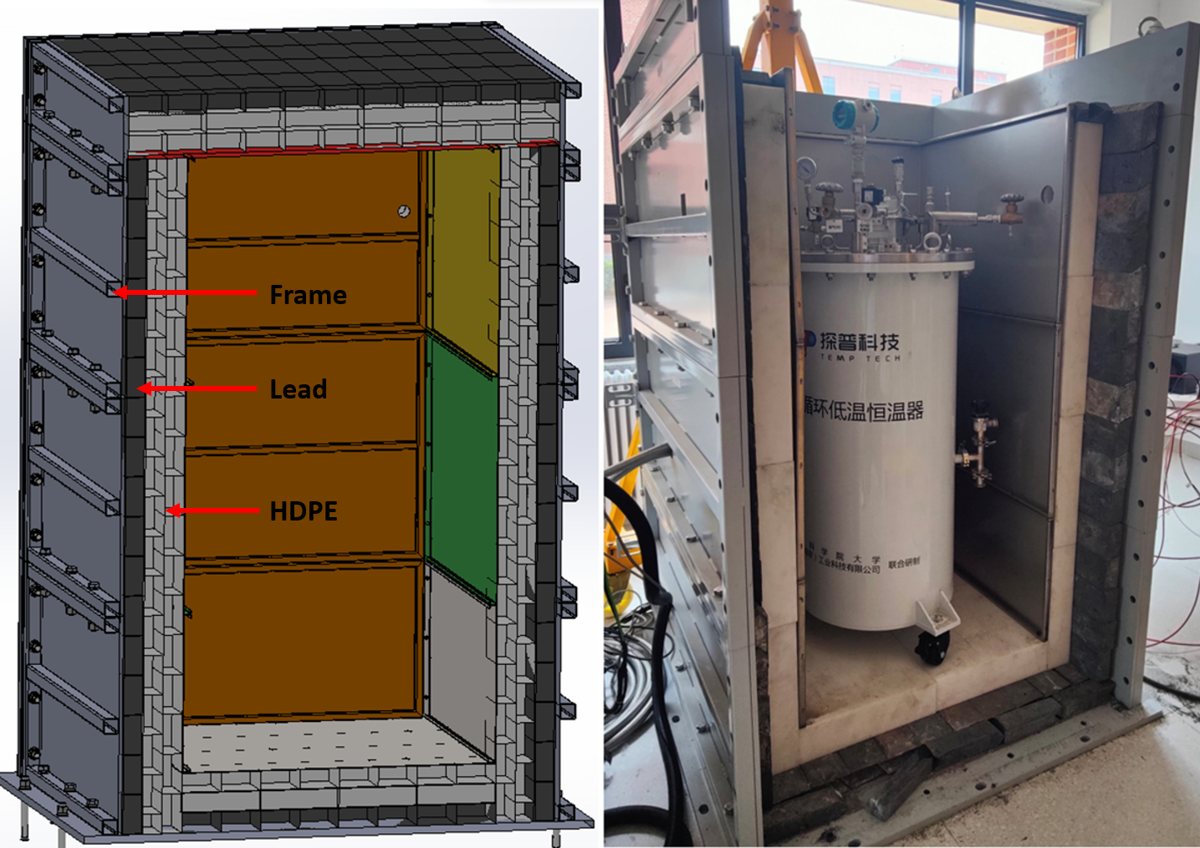}
\caption{Cross-sectional view of the shielding structure (left) and a photograph of the half-assembled shielding with the dewar placed inside (right). The shielding consists of an outer stainless steel support frame, a 5 cm thick lead layer (dark), a 10 cm thick HDPE layer (light), and an inner stainless steel frame.}
\label{fig:shielding}
\end{figure}

All shielding components are mechanically supported by a stainless steel frame, which ensures structural integrity and prevents accidental displacement. To accommodate essential services, the shielding structure includes three dedicated feedthrough holes for signal cables and gas lines, as illustrated in Fig.~\ref{fig:shielding}.

\section{Detector performance}

The detector system was operated continuously for one month under stable cryogenic conditions to evaluate key performance metrics, including light yield, energy resolution, spatial uniformity of the light yield, long-term operational stability, and intrinsic crystal radioactivity. During this period, all critical parameters—such as temperature, PMT gain, and scintillation light yield—were closely monitored to assess the system’s suitability for low-threshold, long-duration rare-event searches.

\subsection{Light yield and energy resolution}

\begin{figure}
  \includegraphics[width=0.5\textwidth]{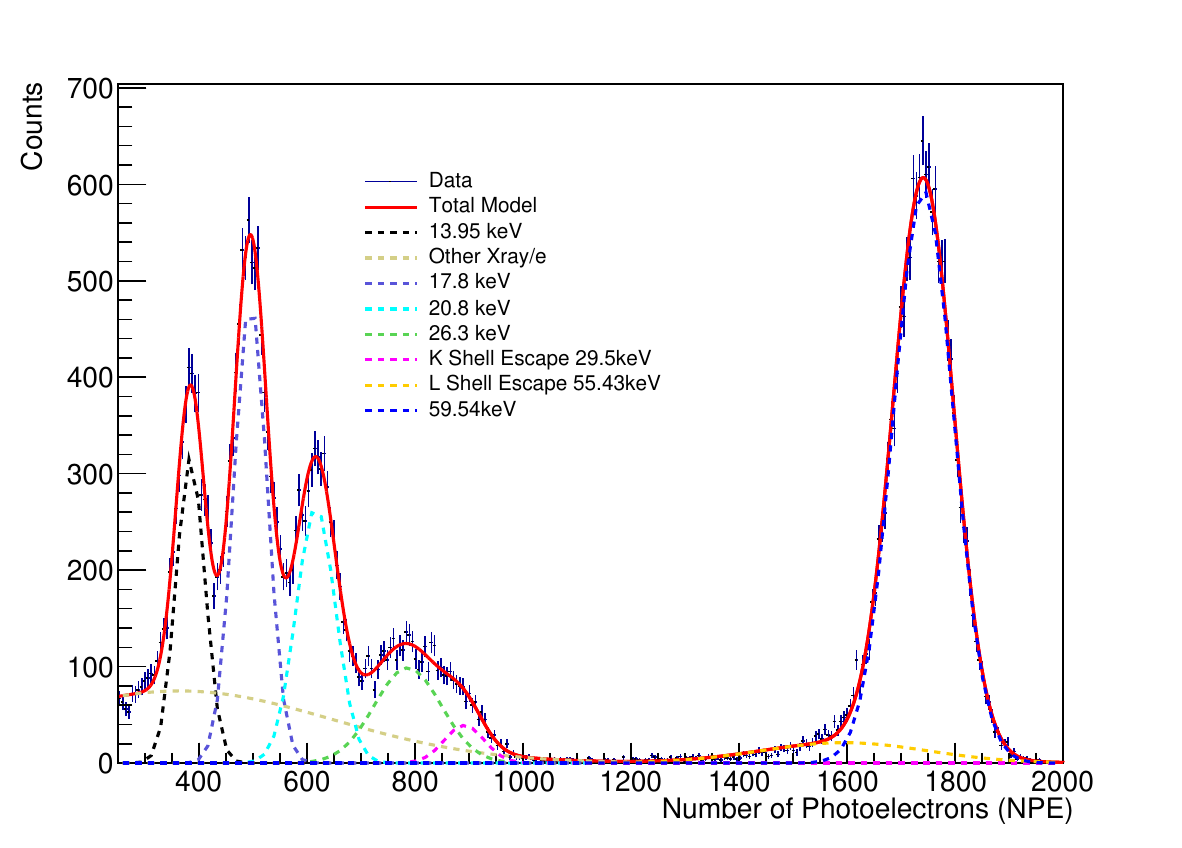}
\caption{Calibration spectra of NO.2 CsI crystal with fitted peaks, recorded after two days of cooling. }
\label{fig:LY}
\end{figure}

The waveform analysis toolkit used in this study is described in detail in a separate publication \cite{LYTemp1Waveform,waveAna}. After determining the average charge of single photoelectrons (SPE), the number of photoelectrons (NPE) produced by a scintillation event is calculated as

\begin{equation}
NPE = {Q_{\mathrm{total}} \over Q_{\mathrm{SPE}}},
\end{equation}

where $Q_{\mathrm{total}}$ is the integrated charge of the waveform, and $Q_{\mathrm{SPE}}$ is the average charge of SPE. The light yield (LY), expressed in photoelectrons per unit energy, is then given by

\begin{equation}
LY\ [\mathrm{PE}/\mathrm{keV}_{ee}] = {NPE \over E{\mathrm{src}}},
\end{equation}

where $E_{\mathrm{src}} = 59.6$~keV for the $^{241}$Am calibration source. The energy resolution is defined as the full width at half maximum (FWHM) of the characteristic peak and is given by 

\begin{equation}
\mathrm{Res} = {2.355 \times \sigma \over E},
\end{equation}

where $\sigma$ is the standard deviation obtained from a Gaussian fit to the photopeak.

\begin{table}
\caption{Light yield and energy resolution of both crystals at approximately 95 K. Measurements were performed with a $^{241}$Am source positioned at the crystal midpoint, using 59.6 keV peak to determine the light yield and different characteristic peaks to determine the energy resolution. }
\label{tab:LYandERes}
\begin{tabular}{lll}
\hline\noalign{\smallskip}
Crystal             & NO.1   & NO.2 \\
\noalign{\smallskip}\hline\noalign{\smallskip}
LY (PE/keV$_{ee}$)  & $29.3 \pm 1.0$   & $28.7 \pm 0.9$  \\
Res. (\%)           &        &       \\
\quad at 13.9 keV   & 15.5   & 19.9  \\
\quad at 17.8 keV   & 15.5   & 14.5  \\
\quad at 20.8 keV   & 15.3   & 13.5  \\
\quad at 59.6 keV   & 7.7    & 7.2   \\

\noalign{\smallskip}\hline
\end{tabular}
\end{table}



Figure~\ref{fig:LY} shows the $^{241}$Am calibration spectrum for both crystals, and the corresponding light yield and energy resolution are summarized in Table~\ref{tab:LYandERes}. The first 3.3 kg crystal achieved a light yield of $29.3 \pm 1.0$ PE/keV$_{ee}$ with a resolution of 7.7 \% (FWHM) at 59.6 keV, while the second yielded $28.7 \pm 0.9$ PE/keV$_{ee}$ with a resolution of 7.2 \%. Such performance is remarkable given the large crystal volume. In general, larger crystals tend to exhibit reduced light yield because scintillation photons must travel longer paths before reaching the photosensor, undergoing more reflections and facing greater chances of absorption or leakage at the boundaries. Through the use of dual-ended readout and enhanced Teflon reflectors, we effectively mitigated these losses and achieved a light yield comparable to much smaller crystals.

For comparison, Jin Liu et al. reported 26.0 PE/keV$_{ee}$ with a 1.0 kg crystal (PMT readout) \cite{LYCompare2}, Lei Wang et al. obtained 30.1 PE/keV$_{ee}$ with a 0.5 kg crystal (SiPM readout) but with 7.8 \% resolution at 59.6 keV \cite{SiPM1}, and Kim et al. measured 26.2 PE/keV$_{ee}$ with a 1 g crystal (SiPM readout) and 21.8 \% at 23.0 keV \cite{LYCompare1}. It is worth noting that all these results were obtained at 77 K, while our measurements were performed at approximately 95 K to ensure PMT stability. Despite the substantially larger crystal mass and the slightly higher operating temperature, the achieved light yield remains competitive with the best reported values, and the energy resolution is superior to most previous measurements, with only one result reaching a comparable level. In our earlier study with a 36.0 g CsI crystal, we carried out a systematic comparison of energy resolutions across various crystal types and experimental setups, where the cubic CsI crystal demonstrated the best performance among both cryogenic pure CsI and doped scintillators~\cite{LYTemp1Waveform}.

\subsection{Light yield spatial uniformity}

\begin{figure}
  \includegraphics[width=0.5\textwidth]{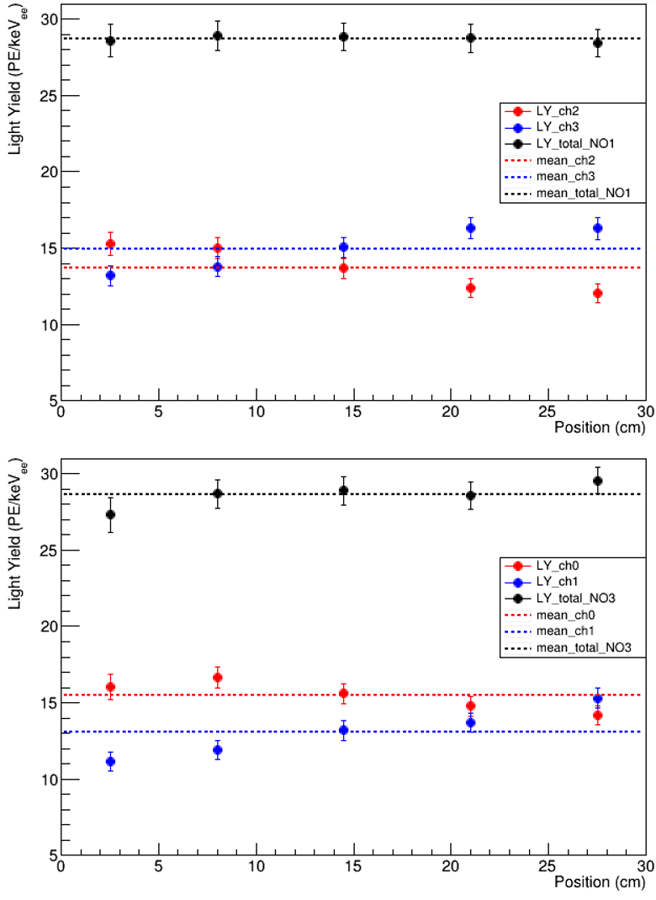}
\caption{Spatial uniformity of light yield along the crystal length. The upper panel shows Crystal No.1 and the lower panel shows Crystal No.2. The observed drops at 27.5 cm (No.1) and 2.5 cm (No.2) occur when the calibration source is near the larger crystal end, where corner light leakage is more pronounced. }
\label{fig:Spatial}
\end{figure}

To evaluate the spatial uniformity of the light yield, the built-in real-time calibration system was employed to position the $^{241}$Am source at various locations along the length of each crystal. The light yield at each position was extracted and fitted using a constant, as shown in Fig.~\ref{fig:Spatial}.

Both crystals exhibit excellent spatial uniformity in light yield. The slight drops observed at 27.5 cm for crystal NO.1 and at 2.5 cm for crystal NO.2 occur when the source is positioned near the larger end of the crystals. At these locations, light leakage from the corners of the oversized output surface becomes more pronounced, resulting in a marginal reduction in collected light. The observed asymmetry in light collection between the two ends reflects the wedge-shaped geometry of the crystals: the larger end typically collects more light, while the smaller end collects less. When the source is placed near one end, that end preferentially collects a larger share of the scintillation photons.

Despite this asymmetry, the total light yield remains stable across all source positions, indicating uniform photon production throughout the crystal volume and efficient light transport to both PMTs.

\begin{figure}
  \includegraphics[width=0.5\textwidth]{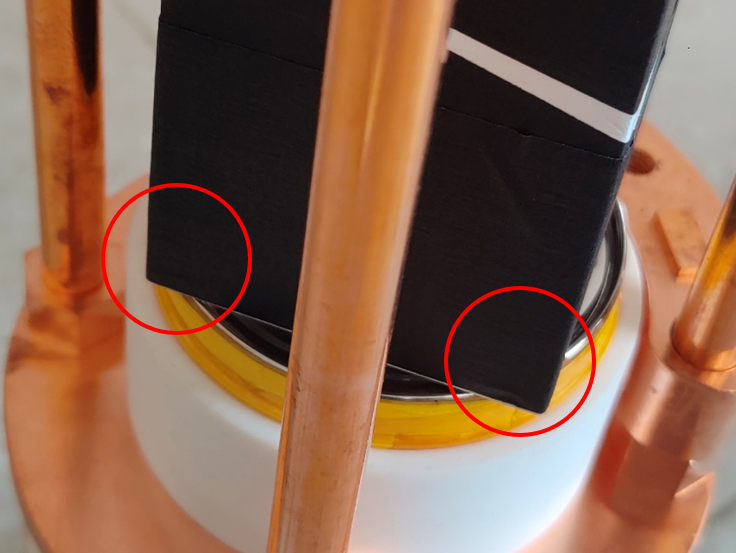}
\caption{The larger end of the wedge-shaped crystal extends beyond the PMT window, potentially causing light leakage from the corners. }
\label{fig:crystalEnd}
\end{figure}

Nevertheless, because the $5 \times 6$ cm$^2$ light output surface is larger than the active window of the PMT (Fig.~\ref{fig:crystalEnd}), a fraction of scintillation light may escape from the edges, leading to a modest reduction in collection efficiency. To mitigate such effects, four layers of Teflon were wrapped around each coupling area. It is anticipated that with improved optical matching between the crystal surface and the PMT window, the overall light yield could exceed 30.0 PE/keV$_{ee}$.

\subsection{Long-term operational stability}

\begin{figure*}
  \includegraphics[width=1.0\textwidth]{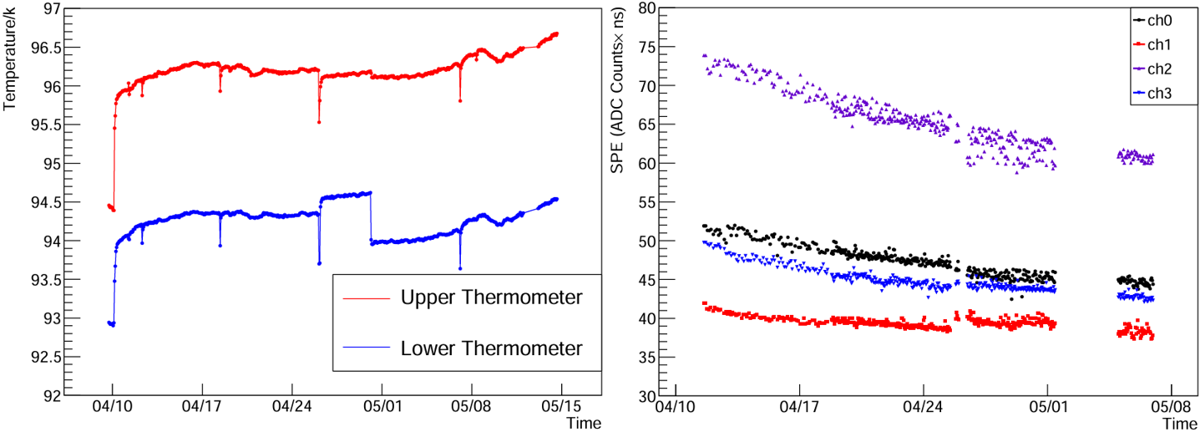}
\caption{Monitoring of temperature (left) and PMT gain (right) during continuous operation. The apparent drops in the temperature curve are caused by vibrations from the motorized calibration system, which would affect the thermal sensor resistance. Gaps in the PMT gain plot correspond to periods when data acquisition was temporarily paused. }
\label{fig:GainAndT}
\end{figure*}

The detector operated reliably under cryogenic conditions for over one month, maintaining a stable temperature of approximately 95 K with minimal fluctuations. The sharp temperature drops observed in the left panel of Fig.~\ref{fig:GainAndT} are attributed to a mechanical interference issue: the thermal sensor wire was inadvertently wound around the drive lever of the calibration module. Movement of the motor during source positioning induced minor vibrations, which temporarily altered the resistance of the thermal sensor, leading to apparent—but non-physical—temperature dips.

\begin{figure}
  \includegraphics[width=0.48\textwidth]{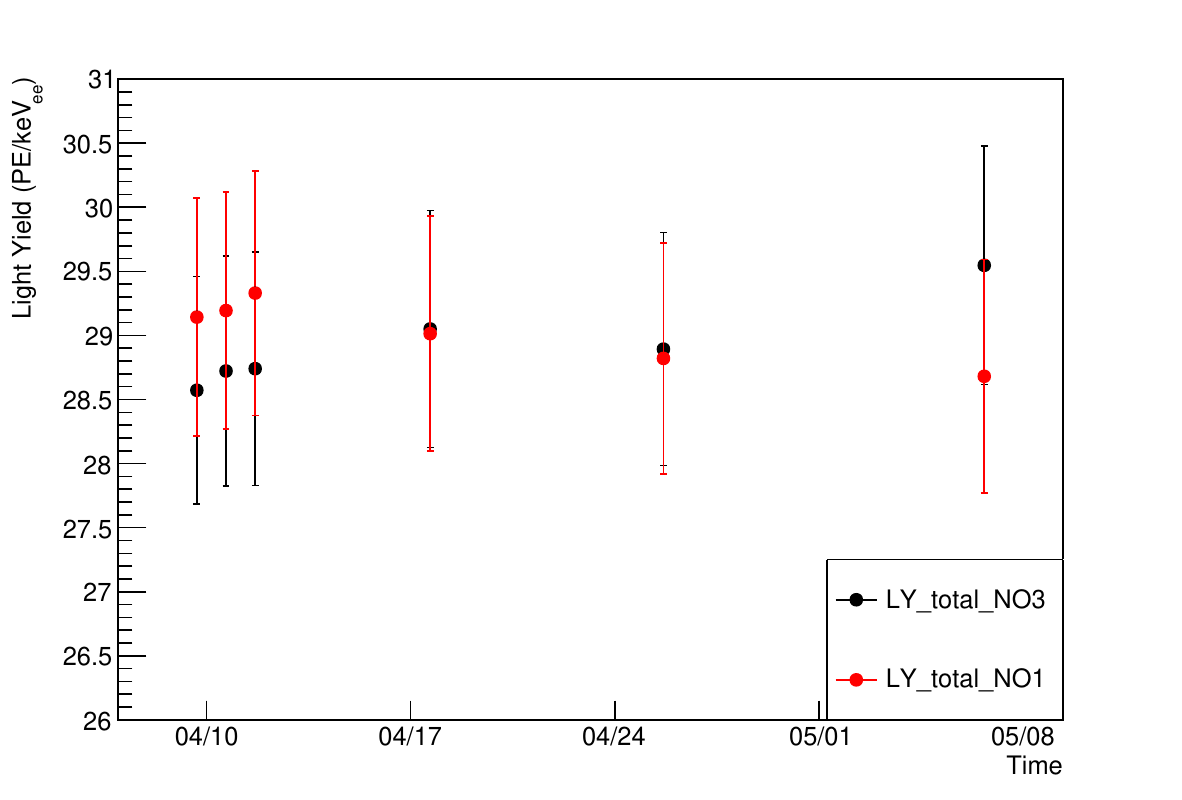}
\caption{Long-term light yield calibration results. All data points were obtained using source calibration positioned at the midpoint of the crystal. }
\label{fig:LYLongTerm}
\end{figure}

Throughout the operation period, both the high voltage and current supplied to the PMTs remained stable. Although a gradual decrease in gain was observed with time, it eventually reached and maintained a stable level, as shown in the right panel of Fig.~\ref{fig:GainAndT}. It is worth noting that several time intervals in the gain plot lack data points; these correspond to periods when data acquisition was deliberately paused, although the detector and cooling system continued uninterrupted.

Regular calibrations with the $^{241}$Am source confirmed consistent light yield performance throughout the extended run. As shown in Fig.~\ref{fig:LYLongTerm}, the measured light yield remained stable at $29.0 \pm 0.1$ PE/keV$_{ee}$ for Crystal No.~1 and $28.9 \pm 0.1$ PE/keV$_{ee}$ for Crystal No.~2 over the entire month-long period.

These results demonstrate, for the first time, that kilogram-scale cryogenic CsI detectors can simultaneously achieve high light yield, excellent energy resolution, and verified long-term stability, establishing their direct applicability to rare-event searches.

\subsection{Self-radioactivity estimation}

\begin{figure}
  \includegraphics[width=0.48\textwidth]{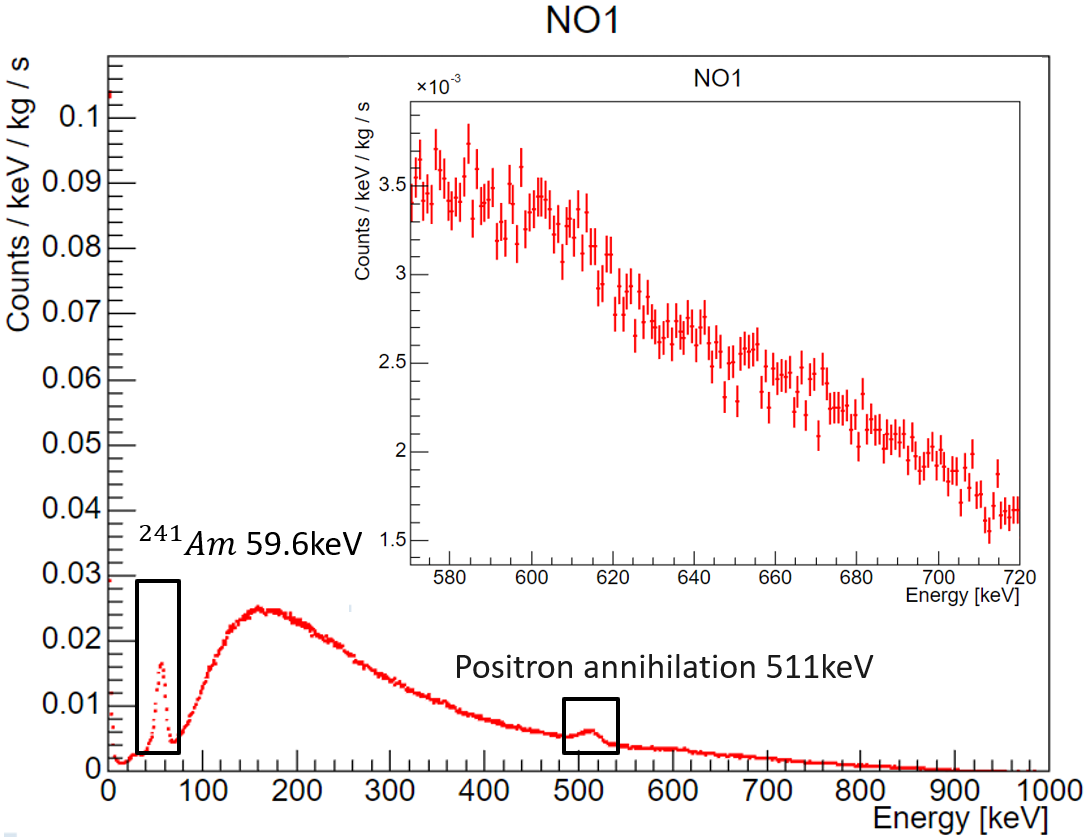}
\caption{Self-triggered energy spectra of NO.1 CsI crystal. The 59.6 keV peak corresponds to the residual characteristic gamma emission of $^{241}$Am, while the 511 keV peak originates from positron–electron annihilation. Characteristic peaks from $^{134}$Cs and $^{137}$Cs are not visible by eye in the spectra. }
\label{fig:SelfTrigESpecNO1}
\end{figure}

To evaluate the intrinsic radioactivity of the CsI crystals, a dedicated 24-hour self-triggered data acquisition run was conducted following the installation of the full 4$\pi$ multi-layer passive shielding. To eliminate interference from the $^{241}$Am calibration source, the source was repositioned at the bottom of the crystal and shielded with a copper block to suppress the emitted gamma rays. This configuration effectively attenuated most of the $^{241}$Am radiation, with only a small residual contribution observed in the low-energy region, as shown in Fig.~\ref{fig:SelfTrigESpecNO1}.

During the measurement, the dewar was fully enclosed within the shielding assembly to minimize external backgrounds, including cosmic muons and ambient radioactivity from the laboratory environment. The characteristic gamma lines of $^{134}$Cs and $^{137}$Cs were used to assess the radiopurity of the crystals \cite{IAEANuclideChart}.

Because the detector’s energy resolution varies with energy \cite{CsIEResolution1,CsIEResolution2}, a resolution parameterization was performed using the standard empirical model \cite{resolutionPara}:

\begin{equation}
\mathrm{Res} = \sqrt{\alpha^2 + {\beta^2 \over E} + {\gamma^2 \over E^2}},
\end{equation}

where $\alpha$ accounts for position-dependent variations in light collection, $\beta$ corresponds to statistical fluctuations in photoelectron production and amplification, and $\gamma$ represents contributions from electronic noise, which could be ignored in practice. The model was therefore simplified to:

\begin{equation}
\mathrm{Res} = \sqrt{\alpha^2 + {\beta^2 \over E}}.
\end{equation}

The 60 keV peak from the $^{241}$Am source and the 511 keV peak from positron-electron annihilation were used to determine the parameters $\alpha$ and $\beta$. The extracted values are listed in Table~\ref{tab:Para}.

\begin{table}
\caption{Parameterization results of energy resolution. Use this model to give the energy region of $^{134}$Cs and $^{137}$Cs characteristic peaks. }
\label{tab:Para}
\begin{tabular}{lll}
\hline\noalign{\smallskip}
Crystal & NO.1    & NO.2 \\
\noalign{\smallskip}\hline\noalign{\smallskip}
$\alpha$            & 0.034   & 0.036    \\
$\beta$             & 0.516   & 0.515    \\
Res at 662 keV      & 3.91 \% & 4.62 \%  \\
Res at 796 keV      & 3.82 \% & 4.07 \%  \\

\noalign{\smallskip}\hline
\end{tabular}
\end{table}

Figure~\ref{fig:SelfTrigESpecNO1} displays the energy spectra obtained from both crystals. No visible peaks are observed near the characteristic gamma-ray energies of $^{134}$Cs and $^{137}$Cs. To quantify potential contamination, the total peak area method \cite{totalPeak}—a widely used technique for estimating the intensity of isolated, non-overlapping peaks—was applied to determine the activity or upper limits ($A_U$) of these isotopes. The results, summarized in Table~\ref{tab:SelfRad}, indicate exceptionally low levels of intrinsic radioactivity, confirming the high radiopurity of the crystals.

\begin{table}
\caption{Radioactivity of CsI crystal. $A_{U}$ gives upper limit. }
\label{tab:SelfRad}
\begin{tabular}{lll}
\hline\noalign{\smallskip}
Activity (mBq/kg) & $^{134}$Cs (796 keV)    & $^{137}$Cs (662 keV) \\
\noalign{\smallskip}\hline\noalign{\smallskip}
NO.1              & $A_{U}=1.67$            & $A_{U}=2.91$  \\
NO.2              & $A_{U}=0.87$            & $A_{U}=2.05$  \\
\noalign{\smallskip}\hline
\end{tabular}
\end{table}

\section{Optical simulation}

To better understand the light collection behavior of the wedge-shaped CsI crystals, optical simulations were performed using the Geant4 toolkit \cite{CEvNS_UCAS}. The simulations aim to reproduce the measured light yield trend as a function of source position and, once validated, provide a cost-effective tool to optimize future crystal geometries without the need for extensive prototyping.

Geant4’s optical model provides a high degree of flexibility, allowing users to define numerous parameters such as surface roughness, reflectivity, absorption length, and scattering characteristics \cite{Geant4Opt1,Geant4Opt2,Geant4Opt3}. While this enables detailed modeling of light transport, it also introduces substantial complexity, making precise quantitative agreement with experimental data difficult and often non-transferable between different experimental configurations.

\begin{figure}
  \includegraphics[width=0.48\textwidth]{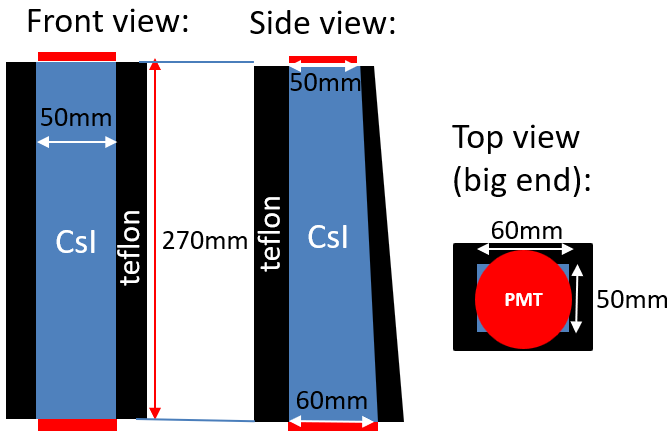}
\caption{Schematic geometry used in the Geant4 optical simulation (not to scale). The blue region denotes the CsI crystal, the black region the Teflon reflector, and the red region the PMT quartz window. A 1 $\mu$m air gap separates the crystal light-output surface from the quartz, where optical photons are collected.}
\label{fig:SimCsI}
\end{figure}

In this study, the optical simulation focused on reproducing the qualitative behavior of the light yield rather than its absolute value or the energy resolution. Specifically, the aim was to reproduce the spatial dependence of light collection and the relative variation between the two readout ends as a function of source position. The simulation geometry, illustrated in Fig.\ref{fig:SimCsI}, emulates the experimental calibration by generating 59.6 keV gamma rays incident vertically into the crystal and tracking the resulting scintillation photons until they reach the air interface. 

\begin{figure}
  \includegraphics[width=0.48\textwidth]{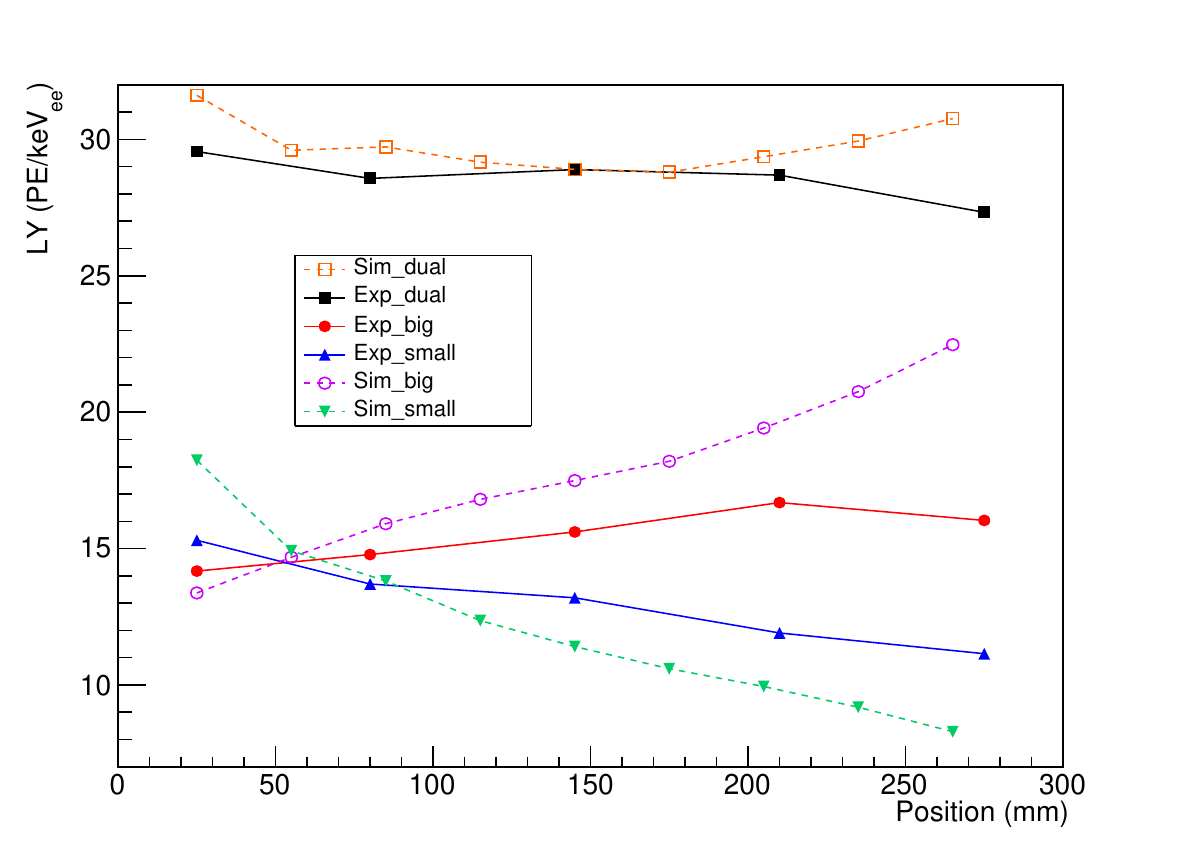}
\caption{Comparison between Geant4 optical simulation results (Sim) and experimental data (Exp), illustrating consistent trends in light collection efficiency as a function of source position. Experimental data from crystal No.2 is shown. }
\label{fig:Simulation}
\end{figure} 

By tuning a limited subset of optical parameters, the simulation successfully reproduced the general trend observed in the experimental data, as shown in Fig.\ref{fig:Simulation}. The simulation results were normalized to the measured light yield for comparison. The optical parameters used in the Geant4 simulation are listed in Table\ref{tab:G4Para}.

\begin{table}[htbp]
\centering
\caption{Geant4 optical surface parameters used in the simulation. Two boundary types are defined: CsI–air and CsI–Teflon. Each boundary is implemented as a two-way interface (e.g., CsI-to-air and air-to-CsI), sharing the same optical properties.}
\label{tab:G4Para}
\begin{tabular}{lll}
\hline\noalign{\smallskip}
Boundary      & Parameter           & Value \\
\noalign{\smallskip}\hline\noalign{\smallskip}
CsI–air       & model               & \texttt{unified} \\
              & finish              & \texttt{polished} \\
              & type                & \texttt{dielectric\_dielectric} \\
              & value               & 0.6 \\
              & reflectivity        & 1.0 \\
              & transmittance       & 0.0 \\
              & efficiency          & 0.0 \\
              & surface roughness   & 0.0 \\
\noalign{\smallskip}
CsI–Teflon    & model               & \texttt{LUT} \\
              & finish              & \texttt{polishedteflonair} \\
              & type                & \texttt{dielectric\_LUT} \\
              & value               & 0.6 \\
              & reflectivity        & 0.98 \\
              & transmittance       & 0.02 \\
              & efficiency          & 0.0 \\
\hline
\end{tabular}
\end{table}

This qualitative trend-matching serves as a validation of the simulation framework, demonstrating its utility as a practical tool for guiding detector design. Although not intended for predictive applications across different experimental platforms, the simulation remains valuable for relative geometry optimization and internal consistency checks.

\section{Conclusion}

We have developed a cryogenic detector system using two 3.3 kg high-purity CsI crystals, achieving a light yield of 29 PE/keV$_{ee}$ and an energy resolution better than 8 \% at 59.6 keV. Despite the large crystal volume, the detector demonstrated excellent spatial uniformity, low intrinsic radioactivity, and stable operation over one month. Optical simulations with Geant4 successfully reproduced the measured light collection trends, supporting future geometry optimization.

The combination of high light yield and excellent energy resolution enables sub-keV detection thresholds and precise recoil spectrum reconstruction, both essential for coherent elastic neutrino-nucleus scattering and low-energy rare-event searches. These results establish cryogenic pure CsI as a viable technology for next-generation neutrino and dark matter experiments. Furthermore, the modular design readily accommodates larger crystals and upgraded readout schemes, ensuring a clear path for future detector advancements.

\begin{acknowledgements}
This work was supported by the National Natural Science Foundation of China (Grant Nos. 12575201, 12175241, 12221005), and by the Fundamental Research Funds for the Central Universities. 
\end{acknowledgements}

\bibliographystyle{spphys}       
\bibliography{reference}   

\begin{thebibliography}{10}
\providecommand{\url}[1]{{#1}}
\providecommand{\urlprefix}{URL }
\expandafter\ifx\csname urlstyle\endcsname\relax
  \providecommand{\doi}[1]{DOI \discretionary{}{}{}#1}\else
  \providecommand{\doi}{DOI \discretionary{}{}{}\begingroup \urlstyle{rm}\Url}\fi

\bibitem{freedman1974coherent}
D.Z. Freedman, Physical Review D \textbf{9}(5), 1389 (1974)

\bibitem{akimov2017COHERENT}
D.~Akimov, J.~Albert, P.~An, C.~Awe, P.~Barbeau, B.~Becker, V.~Belov, A.~Brown, A.~Bolozdynya, B.~Cabrera-Palmer, et~al., Science \textbf{357}(6356), 1123 (2017)

\bibitem{CEvNSFuture}
M.~Abdullah, et~al., arXiv preprint arXiv:2203.07361

\bibitem{weakMixingAngle1}
B.~Ca{\~n}as, E.~Garc{\'e}s, O.~Miranda, A.~Parada, Physics Letters B \textbf{784}, 159 (2018)

\bibitem{weakMixingAngle2ChargeRadius}
M.A. Corona, M.~Cadeddu, N.~Cargioli, F.~Dordei, C.~Giunti, Journal of High Energy Physics \textbf{2024}(5), 1 (2024)

\bibitem{weakMixingAngleNSILightMediator}
V.~De~Romeri, D.K. Papoulias, G.S. Garcia, Physical Review D \textbf{111}(7), 075025 (2025)

\bibitem{formFactor1}
K.~Patton, J.~Engel, G.C. McLaughlin, N.~Schunck, Physical Review C—Nuclear Physics \textbf{86}(2), 024612 (2012)

\bibitem{formFactor2}
E.~Ciuffoli, J.~Evslin, Q.~Fu, J.~Tang, Physical Review D \textbf{97}(11), 113003 (2018)

\bibitem{formFactor3}
K.M. Patton, G.C. McLaughlin, K.~Scholberg, International Journal of Modern Physics E \textbf{22}(06), 1330013 (2013)

\bibitem{magneticMomentNSI}
D.~Papoulias, T.~Kosmas, Physical Review D \textbf{97}(3), 033003 (2018)

\bibitem{magneticMoment}
O.G. Miranda, D.~Papoulias, O.~Sanders, M.~Tortola, J.~Valle, Journal of High Energy Physics \textbf{2021}(12), 1 (2021)

\bibitem{NSI2}
P.B. Denton, Y.~Farzan, I.M. Shoemaker, Journal of High Energy Physics \textbf{2018}(7), 1 (2018)

\bibitem{NSI3}
B.~Dev, K.~Babu, P.~Denton, P.~Machado, C.A. Arg{\"u}elles, J.L. Barrow, S.S. Chatterjee, M.C. Chen, A.~de~Gouv{\^e}a, B.~Dutta, et~al., SciPost Physics Proceedings (2), 001 (2019)

\bibitem{NSI4}
J.~Liao, D.~Marfatia, Physics Letters B \textbf{775}, 54 (2017)

\bibitem{NSI5}
C.~Giunti, Physical Review D \textbf{101}(3), 035039 (2020)

\bibitem{NSI6}
J.~Barranco, O.G. Miranda, T.I. Rashba, Journal of High Energy Physics \textbf{2005}(12), 021 (2005)

\bibitem{lightMediator}
J.B. Dent, B.~Dutta, S.~Liao, J.L. Newstead, L.E. Strigari, J.W. Walker, Physical Review D \textbf{96}(9), 095007 (2017)

\bibitem{lightMediator2}
M.A. Corona, M.~Cadeddu, N.~Cargioli, F.~Dordei, C.~Giunti, Physical Review D \textbf{112}(1), 015007 (2025)

\bibitem{NeutrinoFog1}
B.~Carew, A.R. Caddell, T.N. Maity, C.A. O’Hare, Physical Review D \textbf{109}(8), 083016 (2024)

\bibitem{NeutrinoFog2}
C.A. O’Hare, Physical Review Letters \textbf{127}(25), 251802 (2021)

\bibitem{DarkMatter3}
C.~B{\oe}hm, D.~Cerde{\~n}o, P.~Machado, A.~Olivares-Del~Campo, E.~Perdomo, E.~Reid, Journal of Cosmology and Astroparticle Physics \textbf{2019}(01), 043 (2019)

\bibitem{DarkMatter4}
B.~Dutta, S.~Liao, L.E. Strigari, J.W. Walker, Physics Letters B \textbf{773}, 242 (2017)

\bibitem{DarkMatter5}
T.~Schwemberger, T.T. Yu, Physical Review D \textbf{106}(1), 015002 (2022)

\bibitem{lowerThreshold1}
P.~Barbeau, V.~Belov, I.~Bernardi, C.~Bock, A.~Bolozdynya, R.~Bouabid, J.~Browning, B.~Cabrera-Palmer, E.~Conley, V.~Da~Silva, et~al., Physical Review D \textbf{109}(9), 092005 (2024)

\bibitem{lowerThreshold2}
J.R. Klein, A.~Machado, D.~Schmitz, R.~Strauss, M.~Diwan, C.~Jackson, J.~Maneira, K.~Mavrokoridis, N.~McConkey, T.~Mohai, et~al., Snowmass neutrino frontier nf10 topical group report: Netrino detectors.
\newblock Tech. rep., Pacific Northwest National Laboratory (PNNL), Richland, WA (United States~… (2022)

\bibitem{ReactorNeutrino}
A.C. Hayes, P.~Vogel, Annual Review of Nuclear and Particle Science \textbf{66}(1), 219 (2016)

\bibitem{SolarNeutrino}
Y.~Fukuda, T.~Hayakawa, E.~Ichihara, K.~Inoue, K.~Ishihara, H.~Ishino, Y.~Itow, T.~Kajita, J.~Kameda, S.~Kasuga, et~al., Physical Review Letters \textbf{82}(12), 2430 (1999)

\bibitem{SpectraDistortion1}
J.~Liao, D.~Marfatia, Physics Letters B \textbf{775}, 54 (2017)

\bibitem{SpectraDistortion2}
Y.~Farzan, Physics Letters B \textbf{748}, 311 (2015)

\bibitem{keVnr}
S.~Lee, H.~Joo, H.~Kim, K.~Kim, S.~Kim, Y.~Kim, Y.~Ko, H.~Lee, J.~Lee, H.~Park, et~al., Physical Review C \textbf{110}(1), 014614 (2024)

\bibitem{darkMatter1}
G.~Angloher, A.~Bento, C.~Bucci, L.~Canonica, X.~Defay, A.~Erb, F.~von Feilitzsch, N.F. Iachellini, P.~Gorla, A.~G{\"u}tlein, et~al., The European Physical Journal C \textbf{76}(1), 25 (2016)

\bibitem{darkMatter2}
R.~Essig, G.K. Giovanetti, N.~Kurinsky, D.~McKinsey, K.~Ramanathan, K.~Stifter, T.T. Yu, A.~Aboubrahim, D.~Adams, D.~Alves, et~al., arXiv preprint arXiv:2203.08297  (2022)

\bibitem{SiPM1}
L.~Wang, G.~Li, Z.~Yu, X.~Liang, T.~Wang, F.~Liu, X.~Sun, C.~Guo, X.~Zhang, Y.~Lei, et~al., The European Physical Journal C \textbf{84}(4), 440 (2024)

\bibitem{SiPM2}
K.~Ding, J.~Liu, Y.~Yang, D.~Chernyak, The European Physical Journal C \textbf{82}(4), 344 (2022)

\bibitem{LYTemp1Waveform}
C.G. Su, Q.~Liu, L.Q. Kong, S.~Chen, K.~Moharrami, Y.H. Zheng, J.~Li, Nuclear Science and Techniques \textbf{36}(5), 82 (2025)

\bibitem{LYTemp2}
V.~Mikhailik, V.~Kapustyanyk, V.~Tsybulskyi, V.~Rudyk, H.~Kraus, physica status solidi (b) \textbf{252}(4), 804 (2015)

\bibitem{waveAna}
C.~Su.
\newblock Waveform analysis program (2024).
\newblock \urlprefix\url{https://gitee.com/schg_ucas/wave-ana}

\bibitem{LYCompare2}
D.~Chernyak, D.~Pershey, J.~Liu, K.~Ding, N.~Saunders, T.~Oli, The European Physical Journal C \textbf{80}(6), 547 (2020)

\bibitem{LYCompare1}
W.~Kim, H.~Lee, K.~Kim, Y.~Ko, J.~Jeon, H.~Kim, H.~Lee, arXiv preprint arXiv:2312.07957  (2023)

\bibitem{IAEANuclideChart}
Live chart of nuclides.
\newblock \urlprefix\url{https://www-nds.iaea.org/relnsd/vcharthtml/VChartHTML.html}

\bibitem{CsIEResolution1}
M.~Moszy{\'n}ski, A.~Syntfeld-Ka{\.z}uch, L.~Swiderski, P.~Sibczy{\'n}ski, M.~Grodzicka, T.~Szcz{\k{e}}{\'s}niak, A.~Gektin, P.~Schotanus, N.~Shiran, R.~Williams, IEEE Transactions on Nuclear Science \textbf{63}(2), 459 (2016)

\bibitem{CsIEResolution2}
M.~Moszy{\'n}ski, M.~Balcerzyk, W.~Czarnacki, M.~Kapusta, W.~Klamra, P.~Schotanus, A.~Syntfeld, M.~Szawlowski, V.~Kozlov, Nuclear Instruments and Methods in Physics Research Section A: Accelerators, Spectrometers, Detectors and Associated Equipment \textbf{537}(1-2), 357 (2005)

\bibitem{resolutionPara}
J.~Qin, C.~Lai, B.~Ye, R.~Liu, X.~Zhang, L.~Jiang, Applied Radiation and Isotopes \textbf{104}, 15 (2015)

\bibitem{totalPeak}
G.~Gilmore, D.~Joss, \emph{Practical gamma-ray spectrometry} (John Wiley \& Sons, 2024)

\bibitem{CEvNS_UCAS}
Cevns‑ucas simulation framework (2025).
\newblock \urlprefix\url{https://gitee.com/kong_l_q/cevns_ucas.git}

\bibitem{Geant4Opt1}
G.~Collaboration.
\newblock Physics reference manual, release 10.4 (2017).
\newblock \urlprefix\url{https://indico.cern.ch/event/679723/contributions/2792554/attachments/1559217/2454299/PhysicsReferenceManual.pdf}

\bibitem{Geant4Opt2}
D.J. van~der Laan, D.R. Schaart, M.C. Maas, F.J. Beekman, P.~Bruyndonckx, C.W. van Eijk, Physics in Medicine \& Biology \textbf{55}(6), 1659 (2010)

\bibitem{Geant4Opt3}
E.~Roncali, M.~Stockhoff, S.R. Cherry, Physics in Medicine \& Biology \textbf{62}(12), 4811 (2017)

\end{thebibliography}

%
%

\end{document}